\documentclass[
reprint,
superscriptaddress,
nofootinbib,
 amsmath,amssymb,
 aps,
 prl,
floatfix,
]{revtex4-2}

\usepackage{comment}
\usepackage{graphicx}
\usepackage{dcolumn}
\usepackage{bm}
\usepackage{amsmath}
\usepackage[dvipsnames]{xcolor}

\begin{document}

\preprint{APS/123-QED}

\title{Training and memory in a randomly driven fractal gel}

\author{Chloe W. Lindeman\thanks{These authors contributed equally.}}
\altaffiliation{These authors contributed equally to this work.}
\affiliation{%
Department of Physics and Astronomy, Johns Hopkins University, Baltimore, Maryland 21218, USA
}%

\author{Joshua D. Clugston}
\altaffiliation{These authors contributed equally to this work.}
\affiliation{%
 Department of Physics, University of Ottawa, Ottawa, Ontario K1N 6N5, Canada
}%

\author{Justin C. Goodrich}
\affiliation{%
 National Synchrotron Light Source II, Brookhaven National
 Laboratory, Upton, NY 11973, USA
}

\author{Mark Sutton}
\affiliation{%
 Department of Physics, McGill University, Montréal, Québec H3A 2K6, Canada
}%

\author{James L. Harden}
\email{jharden@uottawa.ca}
\affiliation{%
 Department of Physics, University of Ottawa, Ottawa, Ontario K1N 6N5, Canada
}%
\author{Robert L. Leheny}
\email{leheny@jhu.edu}
\affiliation{%
Department of Physics and Astronomy, Johns Hopkins University, Baltimore, Maryland 21218, USA
}

\date{\today}%
 
\begin{abstract}
A variety of disordered materials, including jammed particulate systems and crumpled paper, can be trained to exhibit memory of a cyclic strain amplitude. Only recently, however, have questions emerged around the whether random driving can impart similar training. Here, we employ x-ray photon correlation spectroscopy to study a fractal nanoparticle gel trained via either deterministic cyclic shear or random shear bounded by strain amplitudes $-\gamma_t$ and $+\gamma_t$.  Both types of training lead to microstructural reversibility, with a slower and more irregular training for the random protocol. In both cases, the shear induces redistribution of internal stress in the gel with corresponding irreversible strain displacements whose magnitudes decrease during training. Finally, we show that memory of the random driving can be read out and quantified using a standard protocol. 
\end{abstract}

\maketitle  

How exactly a disordered solid responds to mechanical deformation is a complicated function of the microscopic configuration, deformation type, and history of the material. It is therefore surprising that, for some materials, the response to repeated driving at a fixed amplitude leaves a clear memory: the amplitude can be unambiguously read out using a protocol that measures particle displacements or configuration changes~\cite{keim2013yielding, keim2014mechanical, fiocco2014encoding, fiocco2015memory, keim2019memory, adhikari2018memory}. Typically, such memories are encoded via the increasing reversibility of the microscopic configuration upon repeated application of cyclic strain. In other words, the microstructure of a well-trained system returns to the same configuration at the end of each cycle (though it may not follow the same path out and back; that is, displacements may be hysteretic). While this behavior is well documented in jammed particulate systems~\cite{mungan2019networks, keim2013yielding, fiocco2014encoding, benson2021memory, royer2015precisely}, crumpled sheets~\cite{shohat2022memory}, and systems engineered to exhibit hysteresis~\cite{bense2021complex, meulblok2026transients,kumar2025self, jules2022delicate, paulsen2026mechanical}, whether it should occur in other disordered materials like gels is not obvious. Moreover, real materials are often subject to driving that is stochastic; the effect of random driving has been explored in recent theoretical work~\cite{mungan2025self, chatterjee2026memory}, but the problem has not been addressed experimentally. In this work, we investigate both of these questions by demonstrating training and memory in a fractal gel subject to cycles of deterministic and random shear strain profiles. 

\begin{figure}
\includegraphics[width=8.6cm]{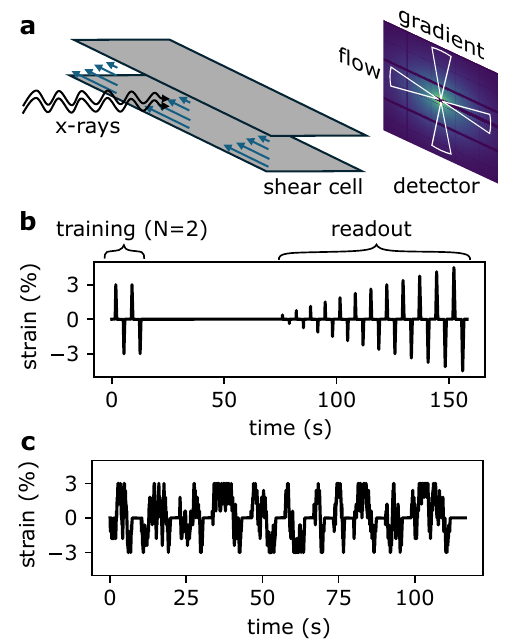}
\caption{(a) Experimental geometry. A coherent x-ray beam is incident on the gel, which is confined within a parallel-plate shear cell.  The incident beam is along the vorticity direction  of the shear; consequently, the scattering wave vectors at the downstream detector lie in the flow-gradient plane. (b) Deterministic training sequence with $\gamma_t = 3\%$ and $N=2$ and corresponding readout protocol. (c) Example random training sequence with $\gamma_t = 3\%$.  }
\label{fig:cartoon}
\end{figure}

Using x-ray photon correlation spectroscopy (XPCS) as depicted schematically in Fig.~\ref{fig:cartoon}(a), we measure changes to a gel's microstructure induced by shear and show that the gel can be trained to exhibit reversible behavior across a range of strain amplitudes. We then demonstrate training with \textit{random} strain cycling, where the strain varies in time as a random walk bounded by values $\gamma_t$ and $-\gamma_t$. Both deterministic and random training exhibit similar behavior as a function of the length scale (wave vector), indicating the microscopic irreversibility is fundamentally the same, although in the random case the training is slower and more irregular. Finally, we use a standard readout protocol to test for memory of the training strain amplitude. For both types of training, we find a signature of the training strain similar to that seen for dense amorphous materials.

\textit{Methods} --- Samples were prepared by slowly mixing vacuum-dried Aerosil 150 (Degussa) into mineral oil (Sigma Aldrich) while stirring to produce a gel with a weight fraction of 6.5\%. 
The XPCS experiments were performed at the CHX beamline of the National Synchrotron Light Source II using 9.65 keV x-rays. The gel was contained in a custom, parallel-plate shear cell with a gap height of 1.5 mm and sample thickness of 1 mm, and a partially coherent x-ray beam was focused to $10 \times 10$ $\mu$m spot on the sample.  The incident beam was along the vorticity axis of the cell as shown schematically in Fig.~\ref{fig:cartoon}(a); consequently, the scattering wave vectors $\vec{q}$, measured at the detector (Eiger 4M) 16.0 m downstream, were in the flow-gradient plane. The intensity $I(\vec{q})$ displayed the character of a fractal gel (see Supplemental Material). The top plate of the cell was held fixed, while the bottom plate was driven by a piezoelectric stage (Physik Instrumente P-628.1). Time-dependent shear waveforms were generated and sent to the stage using a Measurement Computing USB-231 DAQ device with 16-bit analog output resolution, corresponding to nominal displacement steps of $\sim 20$~nm, enabling {\it in situ} application of precise time-varying shear strains to the gel.  

Two types of training experiments were performed: those in which periodic cyclic shear of amplitude $\gamma_t$ was imposed and those in which a random strain sequence was imposed.  The shear rate in all experiments was 0.1 1/s.   For periodic training, the gel was strained via cycles $0 \rightarrow \gamma_{t} \rightarrow 0 \rightarrow -\gamma_{t} \rightarrow 0$, and the strain was held constant for 3 s each time $\gamma=0$ was reached, as shown in Fig.~\ref{fig:cartoon}(b).  The random strain protocols were created by simulating random walks starting at 0 with step size $\Delta \gamma = \gamma_t/5$. One full cycle was defined as in~\cite{mungan2025self}: $\gamma$ must first reach one boundary, then the other, then return to zero. An example random training protocol is shown in Fig.~\ref{fig:cartoon}(c) with 3 s pauses at the end of each full cycle. Note that $-\gamma_t$ may be reached before $\gamma_t$, in which case subsequent cycles are required to be in the same order.  


\textit{Increasing Reversibility} --- We quantify the change in microstructure between two times $t_1$ and $t_2$ via the XPCS instantaneous correlation function, 
\begin{equation}
C(\vec{q}, t_1, t_2) = \frac{\langle I(\vec{q}, t_1) I(\vec{q}, t_2) \rangle}{\langle I(\vec{q}, t_1) \rangle \langle I(\vec{q}, t_2) \rangle}.
\end{equation}
$I(\vec{q}, t)$ is the coherent scattering intensity at $\vec{q}$ and time $t$, and the averages $\langle \cdot \rangle$ are taken over detector pixels corresponding to a small range of wave vectors around $\vec{q}$. Note that, because of the coherent nature of the beam, this correlation compares the exact microstructural configurations at $t_1$ and $t_2$. We focus our analysis on $\vec{q}$ parallel to the gradient direction; hence, the average is taken over a spread of $\pm5^{\circ}$ in this direction and a radial spread of $\Delta |q|=\pm 0.007$ nm$^{-1}$.

\begin{figure*}
\includegraphics[width=17.2cm]{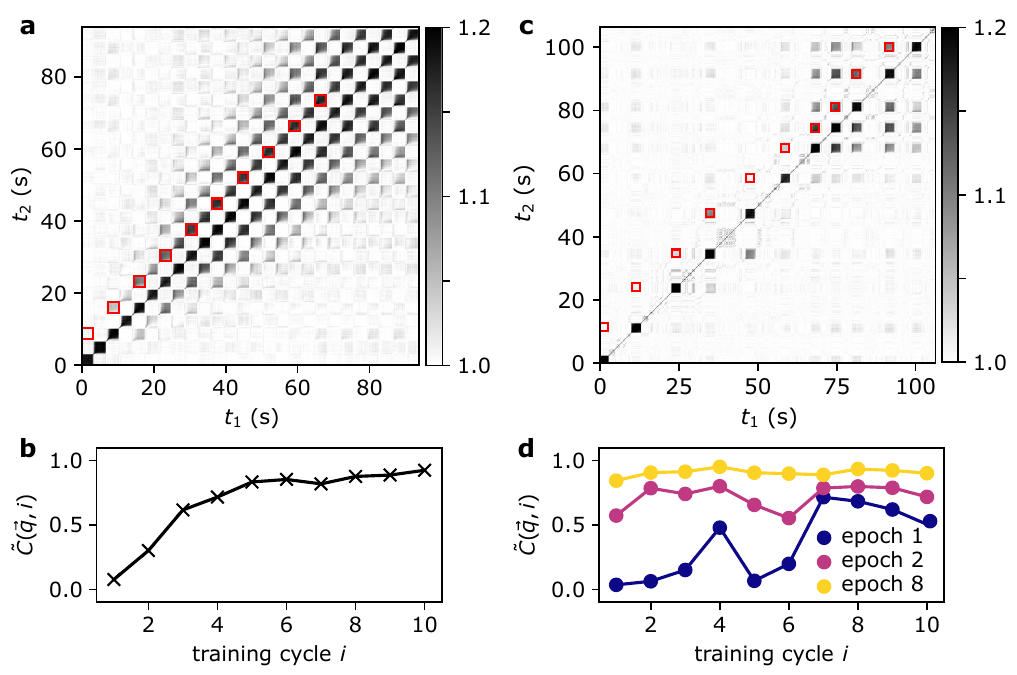}
\caption{(a) Instantaneous correlation function $C(\vec{q},t_1,t_2)$ for deterministic cyclic training and (b) corresponding normalized correlations $\tilde{C}(\vec{q},i)$ that compare the microstructure before and after the $i^{th}$ training cycle. (c) Instantaneous correlation function during the first epoch of random training using the strain profile in Fig~\ref{fig:cartoon}(c); (d) shows the equivalent normalized correlations for multiple epochs. In (a,c), red squares denote times over which $C(\vec{q},t_1,t_2)$ is averaged to obtain $\tilde{C}(\vec{q},i)$ shown in (b,d). All results are for $|\vec{q}| = 0.05$ nm$^{-1}$.}
\label{fig:twotime}
\end{figure*}

Because $C(\vec{q}, t_1, t_2)$ is a function of two times, we visualize it on a two-dimensional heatmap as depicted in Fig.~\ref{fig:twotime}(a) for deterministic training.  Dark regions --- large values of $C(\vec{q}, t_1, t_2)$ --- correspond to values of $t_1$ and $t_2$ for which the microstructures have high correlation; light regions indicate low correlation. The square regions of high correlation along the diagonal correspond to periods at fixed (zero) strain.  Regions of high correlation off the diagonal indicate that the structure has returned partially or in full to a previous configuration; in other words, the displacements during the intervening strain were at least partially reversible. 

We quantify the training by measuring the degree of correlation at times separated by one full cycle by taking the average of $(C(\vec{q}, t_1, t_2)-1)$ at times $t_1$ and $t_2$ when the strain is zero before and after the $i^{th}$ cycle, as outlined by the red squares in Fig.~\ref{fig:twotime}(a), and dividing by $(C(\vec{q}, t_1, t_2)-1)$ during periods of fixed strain (the squares of high correlation along the diagonal) to obtain a normalized correlation $\tilde{C}(\vec{q}, i)$ as a function of training cycle number $i$. Figure~\ref{fig:twotime}(b) shows this correlation for deterministic training with amplitude $\gamma_t = 3\%$.  $\tilde{C}$ increases steadily during the first several cycles, indicating increasing microscopic
reversibility with each cycle of shear.

Note that even as the reversibility grows when comparing across full cycles, the behavior is strongly hysteretic:  comparing across half cycles (eg, comparing the structure at $\gamma=0$ after the application of positive strain with that  at $\gamma=0$ after the application of negative strain) reveals no measurable correlation. This rules out the possibility that the aerosil particles follow the same path out and back as strain is applied and removed as in training of dilute non-Brownian suspensions~\cite{pine2005chaos}; instead, the scattering is consistent with closed-loop trajectories as seen in  jammed systems and elastoplastic models above the so-called hysteresis transition~\cite{keim2013yielding, keim2014mechanical, elgailani2025anomalous}.  
This feature raises the question of whether random strain sequences could train the gel. 

The equivalent instantaneous correlation function and normalized cycle-to-cycle correlations for random training are shown in Fig.~\ref{fig:twotime}(c,d). Because the driving process is random, the length of cycles is longer; hence, the distance between adjacent high-correlation squares on the diagonal is larger.  (Also, pauses at $\gamma = 0$ are included only after full, rather than half, cycles). As in Fig.~\ref{fig:twotime}(a), red boxes indicate regions where $\tilde{C}$ is calculated. In this case, the same 11-cycle random driving protocol was repeated several times; we refer to each repeat as an epoch of training and show $\tilde{C}$ for several epochs in Fig.~\ref{fig:twotime}(d). As with deterministic training (Fig.~\ref{fig:twotime}(b)), the reversibility grows, demonstrating the ability of random strain to train the gel. However, the growth is nonmonotonic, as seen in the data from the first epoch.  Hence, cycles of random strain  can enhance or degrade microscopic reversibility, but the overall trend is toward increasing reversibility. Repeated application of the same random sequence leads to progressively higher reversibility, as seen by the results during subsequent epochs.

\begin{figure}
\includegraphics[width=8.6cm]{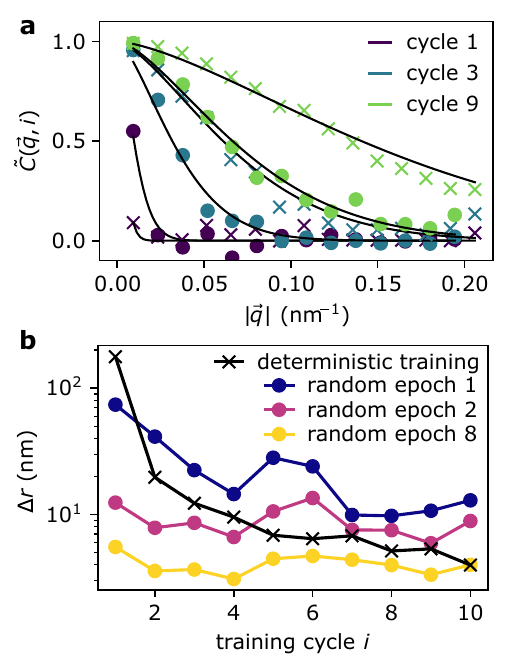}
\caption{(a) Normalized correlation as a function of wavevector magnitude $|\vec{q}|$ for deterministic (x's) and random (o's, all data from first epoch) training cycles with $\gamma_t = 3\%$. Overlaid (solid lines) are the fits from Eq.~(\ref{eqn:compexp}).
(b) Best-fit values of $\Delta r$ for each cycle during several epochs of random training (colorful o's) and during deterministic training (black x's).}
\label{fig:qdep}
\end{figure}

\textit{q-dependence} --- By considering different wave vectors we can examine the scale-dependence of the training. Figure~\ref{fig:qdep}(a) shows $\tilde{C}(\vec{q},i)$ as a function of wave vector magnitude for the trainings depicted in Figs.~\ref{fig:twotime}(a) and (c). Here, each curve indicates the correlations for a different cycle in the training; comparing the untrained configuration with that after the first cycle, we see that the configuration is uncorrelated over all measured length scales; it becomes more correlated faster at lower $q$ (larger length scale), consistent with~\cite{chen2025microstructural}.

We find that the $q$-dependence for both deterministic and random training is well fit by a compressed exponential with exponent 3/2:
\begin{equation}
\tilde{C}(q,i) = \exp\left[-\frac{2\sqrt{3}}{3}(q \Delta r(i))^{3/2}\right].
\label{eqn:compexp}
\end{equation}
This form is equivalent to that observed previously with XPCS, DLS, and simulations of gels and other glassy materials~\cite{cipelletti2000universal, cipelletti2003universal, bouzid2017elastically, chevremont2026tracking} and, as described further in the SM, derives from a model in which the displacements are due to dipolar strain fields centered on local, randomly positioned restructuring events in the gel~\cite{cipelletti2000universal, cipelletti2003universal, bouchaud2001anomalous}.  In this interpretation, $\Delta r(i)$ is the mean irreversible strain displacement (see SM). Figure~\ref{fig:qdep}(b), which shows $\Delta r(i)$ for both deterministic (black x's) and random (colorful o's) driving, quantifies the training process by showing that the displacements decrease as more cycles are applied, with the same non-monotonicity for random training evident in Fig.~\ref{fig:twotime}(d). 
Comparing this non-monotonic trend within each epoch of the random training to the smooth decrease in $\Delta r(i)$ for deterministic training highlights the sensitivity of the \textit{rate} of training to the strain protocol. At the same time, the ability of the same functional form to capture the $q$-dependence of both deterministic and random training is consistent with a shared mechanism. 

\textit{Readout} --- Finally, the development of a reversible response upon training suggests that the gel has stored a memory. We explore this memory using a standard ``readout'' protocol in which a series of strain cycles with the same shape as those in the deterministic training but increasing amplitude, as shown in Fig.~\ref{fig:cartoon}(b), is applied to both deterministically and randomly trained systems that have been trained with different numbers of cycles $N$. In this case, the trained configuration (that is, the structure observed at the start of the readout protocol) is compared with the configuration after each full readout cycle as shown explicitly in SM Fig.~S2(b).   

\begin{figure}
\includegraphics[width=8.6cm]{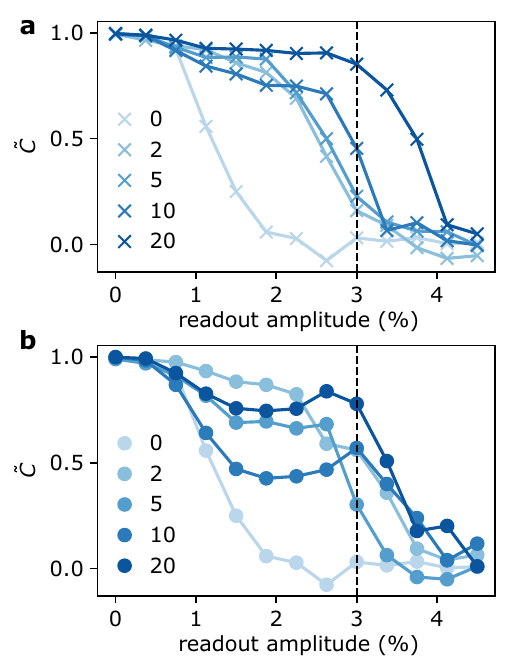}
    \caption{Normalized correlation as a function of readout cycle amplitude for (a) deterministic training and (b) random training after training at $\gamma_t=3$ \% with $N=0$, 2, 5, 10, and 20 cycles. The vertical dashed lines indicate where the reading amplitude reached the training amplitude, $\gamma_t$.}
\label{fig:readout}
\end{figure}

Figures~\ref{fig:readout}(a) and (b) show the resulting readout behavior for systems trained with deterministic and random cycles, respectively. Compared with the untrained systems ($N=0$), for which the correlation rapidly drops as the readout amplitude grows beyond the gel's linear elastic regime, samples that have been trained for even a few cycles develop higher correlation below the training amplitude; that correlation rapidly drops near the training amplitude, the signature of memory of the training amplitude~\cite{fiocco2014encoding, adhikari2018memory, mungan2025self, chatterjee2026memory}. 
The amplitude of $\tilde{C}$ below $\gamma_t$ shows higher variability in the random case, consistent with the irregular, nonmonotonic training under this protocol; however, the overall picture is clear: for both deterministic and random training, cycles of strain strengthen the correlation between adjacent cycles up to the training strain.
To show that this behavior is not a specific to $\gamma_t = 3\%$, we include in the SM similar plots for $\gamma_t  =2\%$ and $4\%$.

\textit{Discussion} --- This work highlights the emergence of amplitude memory in fractal gels. The microstructure becomes reversible whether the training is deterministic, with periodic cycles of strain, or random. Though random training is less efficient, the mechanism appears similar regardless of training type: at the largest length scales the structure quickly exhibits reversibility, while at the smallest scales it may not become fully reversible even after many training cycles. Applying a readout protocol reveals that the gels have not just become reversible but contain a memory of the strain amplitude.

Hysteresis was not a predictable feature of the gel. While other disordered materials have clearly identified microscopic mechanisms for hysteresis --- for example localized particle rearrangements in jammed amorphous solids~\cite{keim2014mechanical, lindeman2025minimal} or individual buckling creases in crumpled paper~\cite{shohat2022memory} --- the network structure of the gel does not present obvious mechanisms for hysteresis. Further, although previous work on concentrated colloidal gels linked memory in those materials to an increase in the local contact number~\cite{schwen2020embedding}, we observe no changes in $I(\vec{q})$ during training to indicate any change in average network structure of the fractal gel (see SM). 
Indeed, the multi-scale response of fractal gels to shear strain~\cite{GislerPRL1999, RajaramSM2010, PergeMannevilleJoR, Colombo_JoR_2014, BrunelJPCB2016, AimePNAS2018, DonleyJoR2022, BantawaNatPhys2023, NabizadehPNAS2024} makes pinpointing specific microscopic processes for hyteresis and training challenging.  However, recent theory that identifies the growth of strain hardening in fractal gels under cyclic shear with local bond reordering (rather than mesoscale restructuring) could serve as a basis for a mechanistic picture of the training and memory formation~\cite{SmithACSNanoTopological}. 

This work raises additional interesting directions for future study. For training amplitudes up to $\gamma_t  =1.0\%$, the gels are fully elastic and reversible; they display neither hysteresis nor transient irreversibility with the onset of shear. However, both hysteresis and a transient process of training are evident at $\gamma_t  =2\%$. Experiments with finer steps in $\gamma_t$ that distinguish whether the two  emerge together or arise independently could provide insight into the microscopic processes involved in training and memory retention in the fractal structure. Finally, the growth in reversibility over multiple epochs of a random sequence suggests the intriguing possibility that the gel learns features of the sequence beyond the strain amplitude. Measurement strategies that can test for this memory capacity would further expand our understanding of the ability of disordered materials to retain signatures of their mechanical history.

\section{Acknowledgments}

We thank Andrei Fluerasu and Lutz Wiegart for technical support and scientific discussions as well as Joey Paulsen and Muhittin Mungan for insightful conversations. JDC and JLH also acknowledge funding from the National Science and Engineering Research Council through Discovery Grant RGPIN-2024-06902.  The research used 11-ID (Coherent Hard X-ray Scattering Beamline) of the National Light Source II; a U.S. DOE Office of Science User Facility operated for the DOE Office of Science by Brookhaven National Laboratory (BNL) under DOE contract DE-SC0012704.   


\bibliography{main}

\end{document}


\renewcommand{\thefigure}{S\arabic{figure}}
\setcounter{figure}{0} 
\renewcommand{\theequation}{S\arabic{equation}}
\setcounter{equation}{0}
\renewcommand{\thetable}{S\arabic{table}}
\setcounter{table}{0}

\title{Supplementary Material for \\ ``Training and memory in a randomly driven fractal gel''}


\maketitle

\renewcommand{\thefigure}{S\arabic{figure}}
\setcounter{figure}{0} 
\renewcommand{\theequation}{S\arabic{equation}}
\setcounter{equation}{0}

\section{Structure of the fumed silica gel}

\begin{figure}
\includegraphics[width=15cm]{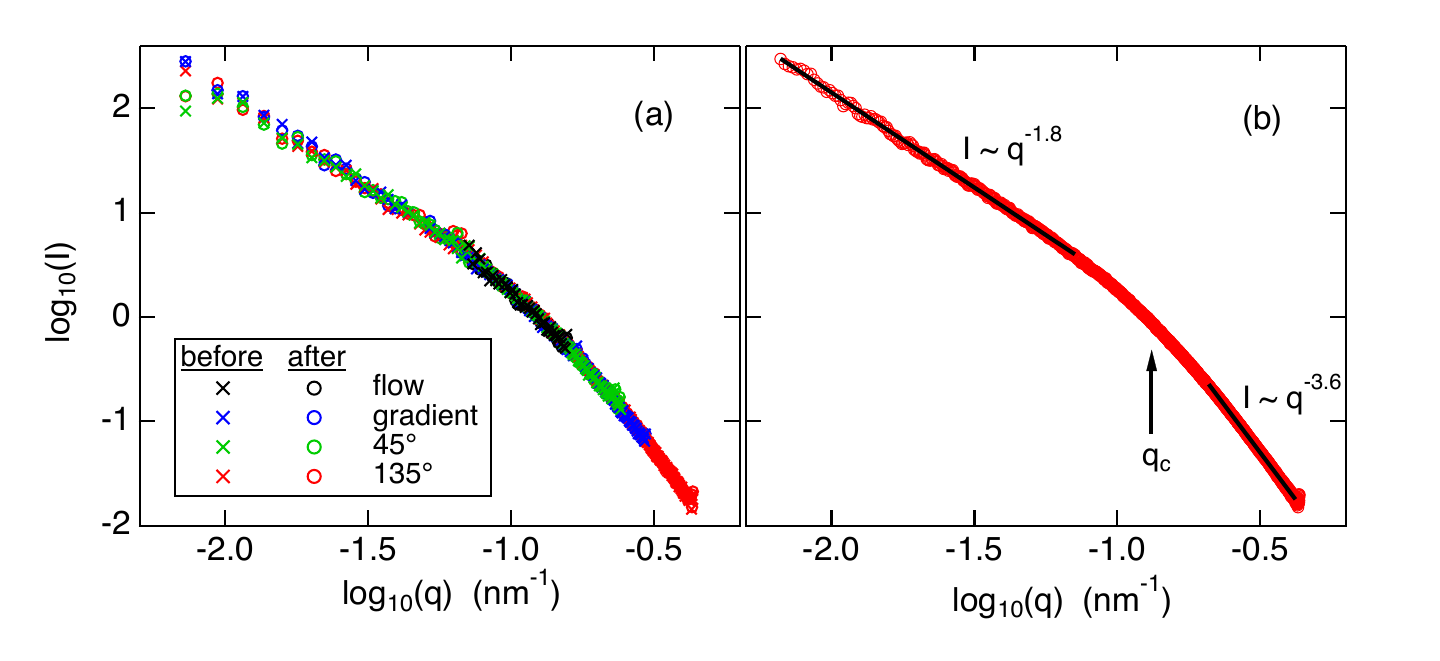}
\caption{(a) SAXS intensity as a function of wave vector on logarithmic scales along four wave-vector directions -- parallel to the shear flow (black), parallel to the shear gradient (blue), at 45$^{\circ}$ to the flow (green), and at 135$^{\circ}$ to the flow (red) --measured before (X's) and after (circles) the training by the random strain sequence depicted in Fig.~1(c) of the main text.  (b) Azimuthally averaged SAXS intensity as a function of wave-vector magnitude on logarithmic scales.  The power-law region at lower $q$ is indicative of fractal network with fractal dimension $D = 1.8$.  The power-law region at higher $q$ is indicative of surface scattering from rough aerosil particle surfaces characterized by a surface fractal dimension $D_s = 2.4$.  The crossover wave vector $q_c$ provides a measure of the lower limiting length scale of the fractal structure, $a \approx 2\pi/q_c \approx 45$ nm.}
\label{fig:Iq}
\end{figure}

As part of the x-ray photon correlation spectroscopy (XPCS) measurements, one obtains the conventional small-angle x-ray scattering (SAXS) intensity $I(\vec{q})$.  Figure \ref{fig:Iq}(a) shows $I(\vec{q})$ measured at the start and end of the experiment in which the gel was subjected to the random strain sequence depicted in Fig.~1(c) of the main text.  Included is the intensity along four wave-vector directions with respect to the shear strain: the flow direction, the gradient direction, at $45^{\circ}$ to the flow, which corresponds to the expansion direction when the strain rate is positive, and at $135^{\circ}$ to the flow, which corresponds to the compression direction when the strain rate is positive.  Over the measured wave-vector range, the intensity is independent of wave-vector direction and is unaffected by the strain sequence.   

Taking advantage of the isotropy in the scattering intensity, for improved statistics we plot in Fig.~\ref{fig:Iq}(b) the azimuthally averaged SAXS intensity $I(q)$.  Over the measurement range, $I(q)$ is comprised of two power-law regions with a crossover near $q_c \approx 0.14$ nm$^{-1}$.  At $q \lesssim q_c$, $I(q) \sim q^{-1.8}$, indicative of a fractal network with fractal dimension $D = 1.8$, which matches the fractal dimension of diffusion-limited cluster aggregation (DLCA).  At $q \gtrsim q_c$, $I(q) \sim q^{-3.6}$. The intensity in this region is consistent with scattering from rough aerosil particle surfaces characterized by a surface fractal dimension $D_s = 6-3.6 = 2.4$.   The crossover wave vector provides a measure of the aerosil particle size, $a \approx 2\pi/q_c \approx 45$ nm. Aerosil 150 has a nominal particle size of 14 nm, so this lower limit suggest the gel is composed of agglomorates of several particles.

\section{Derivation of Wave-Vector Dependence of the Normalized Correlation}

In this section, we provide a derivation of the wave-vector dependence of the normalized correlation $\tilde{C}(\vec{q})$ quoted in Eq.~(2) of the manuscript.  The normalized correlation measured in the XPCS experiment is equivalent to the square of the intermediate scattering function $f(\vec{q})$. 
At delay times corresponding to one cycle, $f(\vec{q})$ can be written as, 
\begin{equation}
    f(\vec{q}) = \sum_{j=1}^{N} \exp(i\vec{q}\cdot\vec{u}_j),
\end{equation}
where the sum is over the particles in the gel within the scattering volume, and $\vec{u}_j$ is the irreversible displacement of the $j^{th}$ particle during the cycle.
Following the analysis of the intermediate scattering function considered previously for a collection of motile scatterers~\cite{Berne}, we convert the sum in $f(\vec{q})$ to an integral,  
\begin{equation}
    f(\vec{q}) = \int  P(\vec{u}) \exp(i\vec{q}\cdot\vec{u}_j)d^3\vec{u},
\end{equation}
where $P(\vec{u})$ is the distribution function of the displacements.
With the coordinates such that $\vec{q}$ is along the $z$-axis, this becomes
\begin{equation}
    f(\vec{q}) = \int_0^{2\pi} d\phi  \int_0^{\pi} \sin\theta d\theta  \int P(\vec{u}) \exp(iqu\cos \theta) u^2 du.
\end{equation}

As described in the manuscript, we identify the sources of irreversible displacements due to a cycle of shear with isolated, microscopic restructuring events that give rise to long-range dipolar strain fields like those observed previously in gels undergoing spontaneous structural relaxation~\cite{cipelletti2000universal,guoJCP2011}. The displacements within a single dipole strain field are anisotropic, but the scattering volume presumably contains many such regions, each with a random orientation, so we can assume $P(\vec{u})$ depends only on the magnitude of $\vec{u}$ and carry through the integrations over angles,
\begin{equation}
    f(\vec{q}) = 4\pi\int_0^{\infty} P(u) \frac{\sin(qu)}{qu} u^2 du
\end{equation}
Defining $W(u)$ as the probability that a particle has a displacement with a magnitude between $u$ and $u+du$, we have $W(u)du = 4\pi u^2P(u)du$, so that 
\begin{equation}
    f(\vec{q}) = \int_0^{\infty} W(u) \frac{\sin(qu)}{qu}  du.
\end{equation}
For a dipolar strain field around a localized restructuring event, the displacement varies with distance from the event as 
\begin{equation}
u(r) \sim r^{-\alpha}
\label{uofr}
\end{equation}
with $\alpha = 2$.
The number of particles $dN$ at a distance from an event between $r$ and $r+dr$ varies as $dN \sim r^2dr$.  Combining this with Eq.~(\ref{uofr}) leads to 
\begin{equation}
    \frac{dN}{du} \sim u^{-(3/\alpha +1)}.
\end{equation}
$dN/du$ in turn is proportional to the displacement probability density $W(u)$.  Hence, the intermediate scattering function can be written as
\begin{equation}
f(\vec{q})= \frac{3}{2}u_0^{3/2}\int_{u_0}^{\infty} u^{-5/2}\frac{\sin(qu)}{qu}\,du,
\label{withcutoff}
\end{equation}
where we set $\alpha = 2$ and introduce a lower cutoff $u_0$ to the displacements.  The lower cutoff results from the assumption that each cycle gives rise to a finite concentration of restructuring events, so that the resulting irreversible strain displacement anywhere in the gel is greater than zero.  (For instance, if the restructuring events that occur each cycle give rise to dipolar stress fields of a similar, characteristic strength, then one can relate $u_0$ to the density of events $n$ as $u_0 \sim n^{-2/3}$.)  The prefactor $\frac{3}{2}u_0^{3/2}$ in Eq.~(\ref{withcutoff}) comes from the normalization of $W(u)$ with the cutoff.  That is, if 
\begin{equation}
W(u) = Cu^{-5/2} \text{  for } u>u_0,
\label{Wnormalize}
\end{equation}
then normalization of $W(u)$ requires
\begin{equation}
C=\frac32 u_0^{3/2}.
\end{equation}

With the change in variables $s=qu$, the intermediate scattering function becomes
\begin{equation}
f(\vec{q})
=
\frac32 (q u_0)^{3/2}
\int_{q u_0}^{\infty}
s^{-7/2}\sin s\,ds.
\end{equation}
This integral can be re-expressed in terms of the Gamma function or evaluated numerically.  Alternatively, one can show that for small $qu_0$, it can be approximated as 
\begin{equation}
f(\vec{q})
\approx
\exp\left[-(qu_0)^{3/2}\right].
\label{compressedexpon}
\end{equation}
As mentioned above, the normalized correlation $\tilde{C}(\vec{q})$ is equivalent to the square of $f(\vec{q})$.  Further, one can show that for the displacement probability distribution given by Eq.~(\ref{Wnormalize}), the mean displacement $\Delta r$ is related to the lower cutoff by $\Delta r = 3u_0$.  Hence, one has
\begin{equation}
\tilde{C}(\vec{q}) = 
\exp\left[-\frac{2\sqrt{3}}{9}(q\Delta r)^{3/2}\right].
\end{equation}
Finally, we note that although Eq.~(\ref{compressedexpon}) is obtained for small $qu_0$, not all of the results for $\tilde{C}(\vec{q})$ in Fig.~3 of the manuscript fall in the range $qu_0 <1$.  However, the data for which $\tilde{C}(\vec{q})$ decays sufficiently slowly as a function of $q$ to assess its shape accurately do fall in this range, and we take the good quality of the fits for these cases as adequate agreement to employ the lineshape.

\section{Memory Readout Following different Training Strain Amplitudes}

\begin{figure*}
\includegraphics[width=17.2cm]{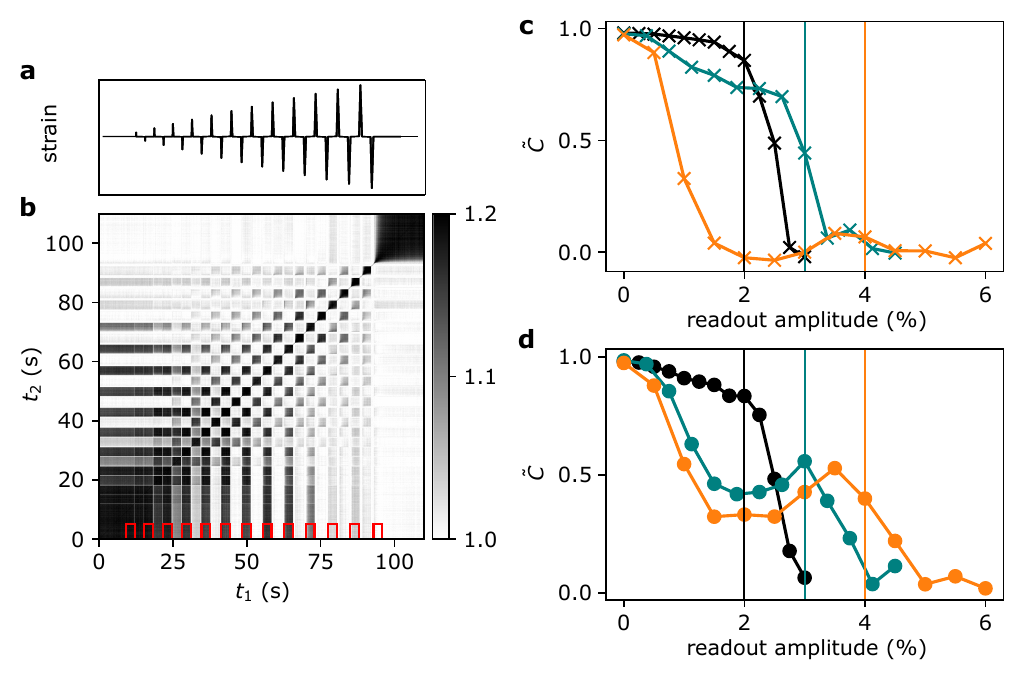}
\caption{(a) Strain protocol during a memory readout experiment and (b) instantaneous correlation function during the same readout for a system trained deterministically with $N=10$ and $\gamma_t = 3\%$. Regions used to calculate values of the normalized correlations for tests of memory are outlined in red. Readout for (c) deterministic training and (d) random training conducted at different training amplitudes (shown in their corresponding color by vertical lines). In all cases, the number of training cycles $N=10$.}
\label{fig:readout-ampl}
\end{figure*}

Figure~\ref{fig:readout-ampl}(a) shows the instantaneous correlation function $C(\vec{q},t_1,t_2)$ measured during a memory readout experiment following deterministic training with $N=10$ and $\gamma_t = 3$.  The regions outlined in red denote the times over the average of ($C(\vec{q},t_1,t_2)$-1) was taken and divided by ($C(\vec{q},t_1,t_2)$-1) during periods of fixed strain (the squares of high correlation along the diagonal) to obtain normalized correlations to  test for amplitude memory. Results for the normalized correlations from readout experiments with different training amplitudes $\gamma_t$ are shown in Figs.~\ref{fig:readout-ampl}(c) and (d) for deterministic and random training, respectively. While readout after training at $\gamma_t = 2\%$ and $\gamma_t=3\%$ provides a clear signature of the training strain amplitude for both deterministic and random training, at $\gamma_t=4\%$ neither case shows a clear signature in the readout. This suggests either that $N=10$ cycles was not enough to train the system at $\gamma = 4\%$ or that $4\%$ strain is near the upper limit of the memory retention in the gel.

Note that readouts for the  randomly trained gels appear to have a slightly different character compared with those from deterministically trained gels, with a peak in the correlation at or near the training strain. Further studies are needed to determine whether this is a generic feature of the randomly trained samples. 

\bibliography{main}